%% file: test.tex
\documentclass[conference, compsoc, letterpaper]{IEEEtran}

\usepackage{microtype}
\usepackage{tikz}
\usepackage{tabularx}
\usepackage{amsmath}
\usepackage{wasysym}
\usepackage{fontawesome}
\usepackage{float}
\usepackage{xspace}
\usepackage{hyperref}
\usepackage[capitalize,noabbrev,nameinlink]{cleveref}
\usepackage[framemethod=tikz]{mdframed}
\usepackage{booktabs}
\usepackage{multirow}
\usepackage{multicol}
\usepackage{listings}

\usepackage{colortbl}

\newcommand{\xyes}{$\CIRCLE$}
\newcommand{\xhalf}{$\LEFTcircle$}
\newcommand{\xno}{$\Circle$}

\title{ChatGPT, is this real? The influence of generative AI on writing style in top-tier cybersecurity papers.}

\author{
    \IEEEauthorblockN{Daan Vansteenhuyse}
\IEEEauthorblockN{
  \textit{DistriNet, KU Leuven} \\
3001 Leuven, Belgium \\
daan.vansteenhuyse@kuleuven.be}
}

\begin{document}
\maketitle

\begin{abstract}
With the release of ChatGPT in 2022, generative AI has significantly lowered the cost of polishing and rewriting text.
Due to its widespread usage, conference organizers instated specific requirements researchers need to adhere to when using GenAI.
When asked to rewrite text, GenAI can introduce stylistic changes, often concentrated to a handful of ``marker words`` commonly associated with AI usage.
Prior large-scale studies in preprints and biomedical science report post-2022 discontinuities of those marker words and broad linguistic features.

This paper investigates whether similar patterns appear in top-tier cybersecurity conference papers (NDSS, USENIX Security, IEEE S\&P, and ACM CCS) over the period 2000-2025. Using text extracted from paper PDFs, we compute lexical and syntactic metrics and track curated marker-word usage. Our findings reveal a gradual long-run drift toward higher lexical complexity and a pronounced post-2022 increase in marker-word usage across all venues showing an emerging trend towards more complex language in cybersecurity papers possibly hindering accessibility.

\end{abstract}

\section{Introduction}
The presentation of research results is as critical as the results themselves. Research characterized by high technical merit, but poor writing quality can lead to reviewer confusion and potential rejection. Consequently, the academic community emphasizes the development of writing skills for junior researchers, often through external training and faculty mentorship \cite{Zhu2004Faculty}. However, the public release of ChatGPT in late 2022 introduced a paradigm shift: LLMs are now widely accessible as sophisticated writing aids.

These models can rapidly polish sentences, smooth transitions, and rewrite entire paragraphs into a preferred academic style. Because this can fundamentally alter a paper's overall stylistic signature, conference organizers have begun implementing specific requirements regarding LLM usage and disclosure. While conferences typically mandate that AI-generated ideas or text must be declared, using an LLM as an advanced spellchecker often does not require explicit mention. 

Nevertheless, even when used solely for polishing, LLMs can introduce measurable stylistic shifts. Recent work on preprint corpora has identified specific linguistic markers that have increased significantly since 2022 \cite{Kobak2025ExcessVocab,Liang2025QuantifyingLLMUsage,GengTrotta2024ChatGPTStyle,Bao2025LinguisticShifts}.

\textbf{Scope}
This paper explores how generative AI has influenced the cybersecurity research community. We conduct a longitudinal study of 25 years of papers published in the four leading (A*) cybersecurity conferences to observe stylistic changes following the adoption of LLMs. By analyzing current Calls for Papers (CFPs), we also identify how GenAI policies vary between venues.
We find that after the introduction of ChatGPT (November 2022), the mean length of words and the frequency of longer words slightly increases.
Additionally, words such as \texttt{delve} and \texttt{enhancing} show a steep increase in usage, likely to be attributed to the usage of GenAI.
We end our work with a discussion of important insights and highlight the influence generative AI can have on papers.

\textbf{Research questions}
We structure this study around three research questions:
\begin{itemize}
  \item \textbf{RQ1 (Conference policy):} What policies do conferences have regarding the usage of GenAI?
  \item \textbf{RQ2 (Stylistic changes):} How has the writing style in A* cybersecurity papers evolved from 2000 to 2025?
  \item \textbf{RQ3 (Words usage):} Are specific words used more frequently following the widespread adoption of LLMs?
\end{itemize}

\input{policytable}
\section{Background and related work}
Meta-research within the security and privacy community has traditionally examined artifacts, evaluation processes, and reproducibility practices \cite{winter_retrospective_2022,olszewski_get_2023,klemmer_how_2025,childers_artifact_2017}. These studies show that community-wide policies and incentives, such as artifact evaluation committees, can measurably change research behavior.

In parallel, emerging work outside of cybersecurity reports that writing in the LLM era exhibits distinct lexical and stylistic shifts.
Bao et al. for example examined preprints published in arXiv and studied linguistic metrics such as readability and complexity, finding an increase in complexity and a decrease in readability while also noting an increase of the usage of certain words \cite{Bao2025LinguisticShifts}.

Complementary, Kobak et al. found a similar increase of certain words in biomedical publications while Geng and Trotta did the same for arXiv papers \cite{GengTrotta2024ChatGPTStyle}

Where our study focuses on general stylistic trends, other work also tried to use the found trends to train neural networks on identifying which papers were written using GenAI \cite{Wang2023SeqXGPT,Liang2025QuantifyingLLMUsage}.

\section{Conference Policies}
\label{sec:genai-policies}
To answer \textbf{RQ1}, we summarize the CFPs of the A* cybersecurity conferences \cite{acm_sigsac_acm_2025,usenix_usenix_2025,internet_society_previous_2025,ieee_ieee_2025}. Table \ref{tab:policy} provides an overview of these policies across four dimensions: author responsibility, disclosure requirements, mechanical assistance, and environmental reporting.

Although ChatGPT was released in November 2022, it took until their 2025 edition for NDSS to formalize a policy.
In 2026, the remaining A* venues followed making them relatively late to adopt a policy.

All venues explicitly state that authors bear full responsibility for the accuracy and integrity of content assisted by GenAI. This reinforces the norm that AI is a tool rather than a co-author. However, venues differ on disclosure: CCS and IEEE S\&P require structured disclosure, while USENIX Security does not require a dedicated section. Most venues permit using AI for grammar and spell-checking without disclosure, except for IEEE S\&P, which requires disclosure even for limited assistance. Notably, IEEE S\&P is the only venue that encourages authors to report the environmental footprint of their GenAI usage using online tools \cite{lacoste2019quantifying}.
\begin{figure*}[th]
  \centering
  \includegraphics[width=\textwidth]{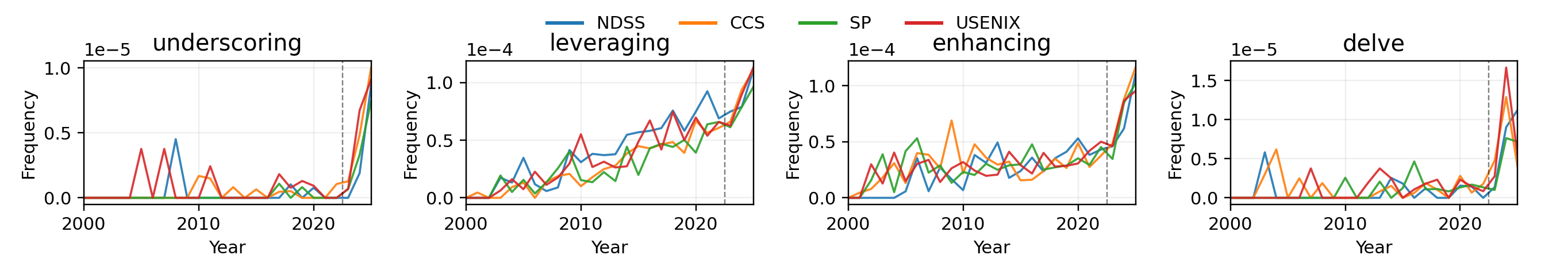}
  \caption{Frequency trends for four words likely influenced by GenAI, showing sharp increases after 2022.}
  \label{fig:markerword}
\end{figure*}

\section{Language Trends}
As generative AI is adopted more widely and conference policies are introduced, it is crucial to determine if these shifts are influencing the style of security papers. Our analysis is based on a dataset comprising all papers published in NDSS, USENIX Security, IEEE S\&P, and ACM CCS from 2000 through 2025. 

To analyze this dataset, we extracted text from PDFs and applied some preprocessing. This included normalizing whitespace, removing detectable page headers and footers, and collapsing hyphenated line breaks caused by column formatting. To ensure the analysis focused on original narrative text, we automatically identified and removed reference sections based on section headings and layout cues. Within this cleaned dataset, we examined established style metrics and tracked the frequency of words commonly associated with AI.

\subsection{Style metrics}
To answer \textbf{RQ2}, we focused on three metrics typically associated with GenAI-influenced writing \cite{Bao2025LinguisticShifts}: long-word rate, mean word length, and Flesch Reading Ease.

\begin{figure}[H]
  \centering
  \includegraphics[width=0.48\textwidth]{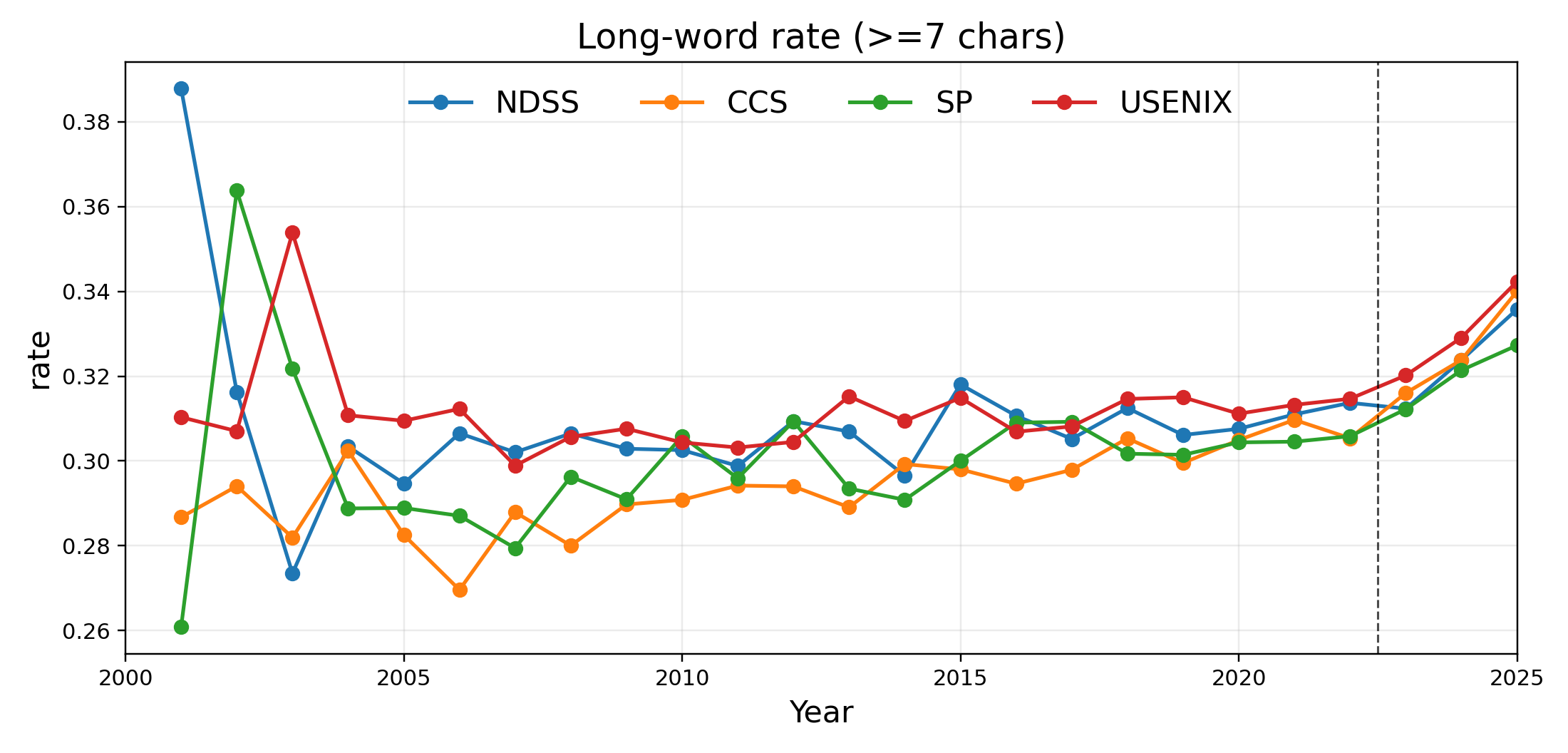}
  \caption{The average percentage of words in a paper consisting of at least 7 characters across all venues.}
  \label{fig:longwordrate}
\end{figure}

\subsubsection{Long-word rate}
The long-word rate serves as a first indicator for lexical complexity and formality. We calculated this as the fraction of words with a length of at least seven characters. As shown in Figure~\ref{fig:longwordrate}, these rates were relatively stable throughout the 2010s. However, we observe a modest, venue-wide uptick starting in 2022, with an upward trend continuing into 2025. Since the historical trend has been stable, it is highly likely that this shift is due to increased LLM usage, as AI-generated text often tends toward longer, more ``academic-sounding'' vocabulary to convey formality~\cite{Bao2025LinguisticShifts}.

\begin{figure}[H]
  \centering
  \includegraphics[width=0.48\textwidth]{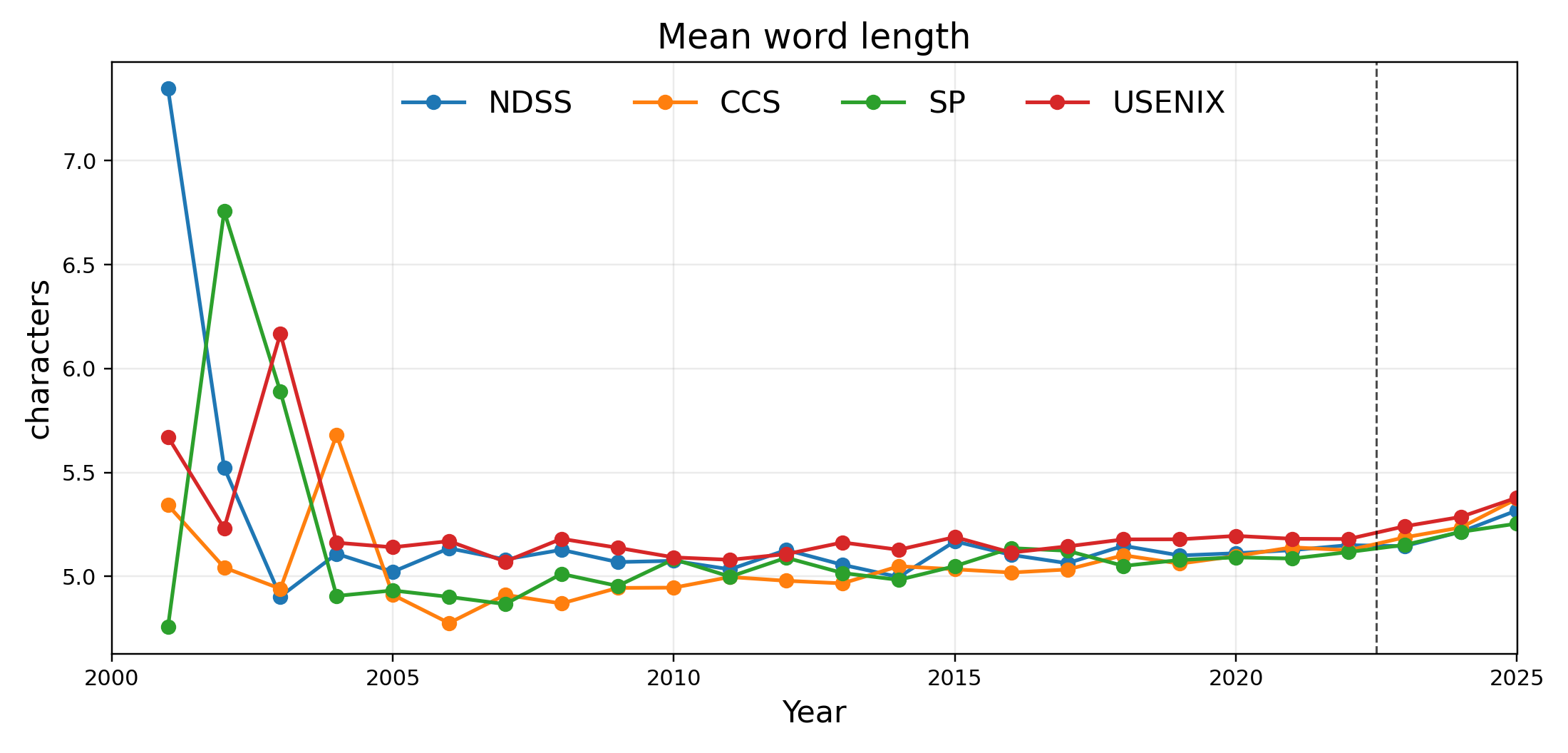}
  \caption{The mean length of words averaged over all papers by year, showing an increase after 2022.}
  \label{fig:meanwordlength}
\end{figure}

\subsubsection{Mean word length}
Mean word length summarizes the average number of characters per token. We calculated the arithmetic mean of token lengths across each document. Mirroring the increase in long words, the mean word length also shows a slight but consistent increase starting in 2023, as illustrated in Figure \ref{fig:meanwordlength}.

\begin{figure}[H]
  \centering
  \includegraphics[width=0.48\textwidth]{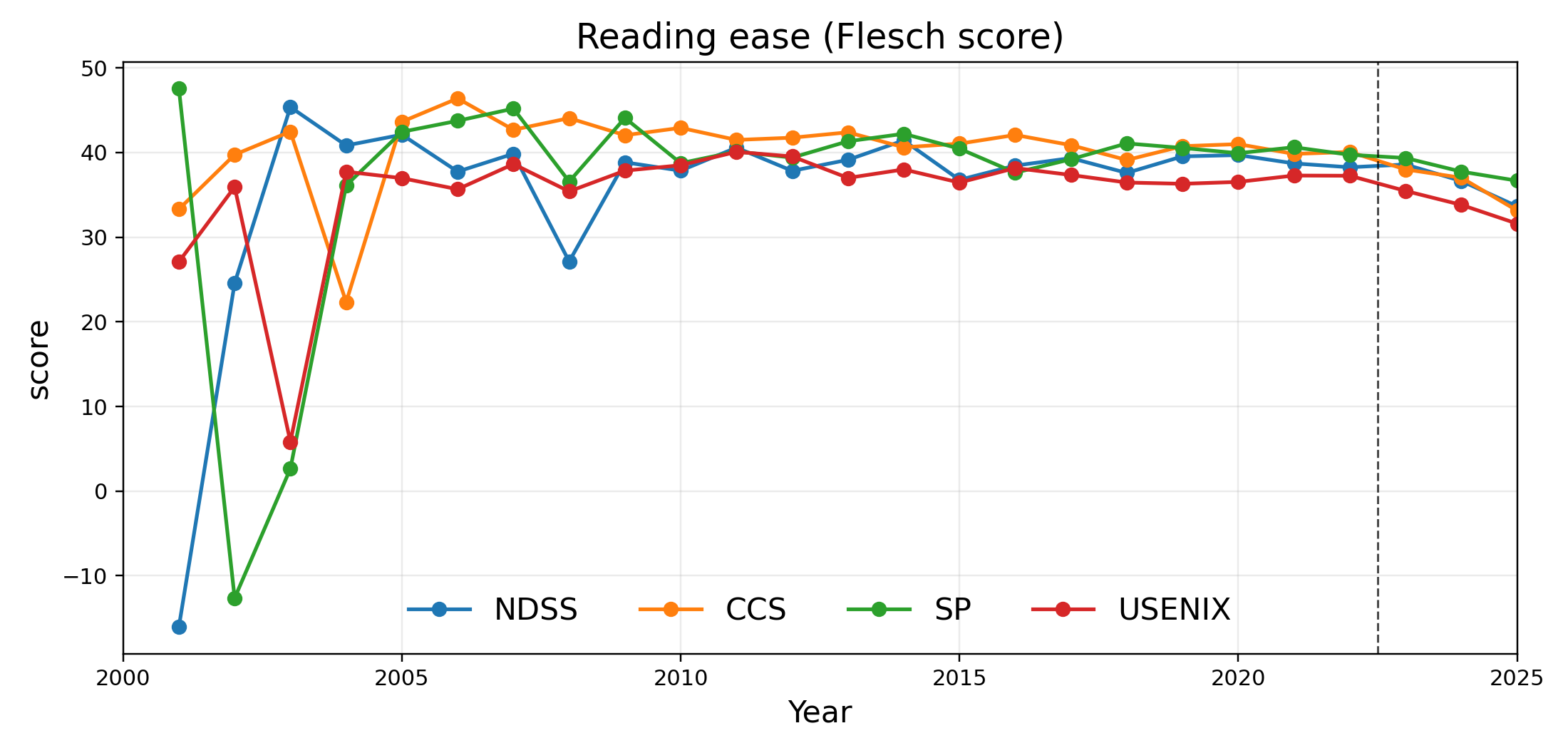}
  \caption{The average Flesch reading ease for papers by year, showing a decrease after 2022.}
  \label{fig:readingease}
\end{figure}

\subsubsection{Reading Ease}
Reading ease assesses text accessibility using the Flesch–Kincaid readability test \cite{farr1951simplification}. This formula combines words per sentence and syllables per word.
We compute Flesch Reading Ease (FRE) as:
\[
  \mathrm{FRE} = 206.835 \;-\; 1.015 \cdot \frac{\text{words}}{\text{sentences}} \;-\; 84.6 \cdot \frac{\text{syllables}}{\text{words}}.
\]
A higher value means the text is easier to read while lower values are harder. Typical guideposts are: \(\approx 90\text{--}100\) (very easy, simple prose), \(60\text{--}70\) (plain English), \(30\text{--}50\) (difficult/technical prose), and \(0\text{--}30\) (very difficult, advanced academic writing) \cite{farr1951simplification}.
  
As shown in Figure \ref{fig:readingease}, all venues settled into a stable band during the 2010s but have exhibited a gradual decline in reading ease since 2022. This decline is consistent with the increase in longer, more complex words. While LLMs can be prompted to simplify text, academic authors often leverage them for more formal phrasing, which typically results in lower readability scores.

\section{Words Usage}
\begin{table}[t]
\caption{Average frequency across A* venues for selected marker words at five reference years.}
\label{tab:markerword_avgs}
\centering
\begin{tabular}{lrrrrr}
\toprule
\textbf{Word (exponent)} & \textbf{2005} & \textbf{2010} & \textbf{2015} & \textbf{2020} & \textbf{2025} \\
\midrule
\texttt{underscoring} \,(\(\times 10^{-6}\)) & 3.7  & 1.6 & 0.6 & 0.8 & 8.8 \\
\texttt{leveraging} \,(\(\times 10^{-5}\))   & 1.7  & 2.7 & 4.2 & 6.2 & 10.0 \\
\texttt{enhancing} \,(\(\times 10^{-5}\))    & 1.8  & 2.0 & 2.4 & 4.1 & 10.0 \\
\texttt{delve} \,(\(\times 10^{-6}\))        & --   & 2.5 & 1.5 & 2.0 & 6.9 \\
\bottomrule
\end{tabular}
\end{table}

Using words identified in existing literature as markers of GenAI usage \cite{Bao2025LinguisticShifts}, we analyzed their frequency in our dataset to address \textbf{RQ3}.
For each conference edition, we count their frequency as the amount of times the word is used divided by the total words across all papers for that edition.
The results can be found in table \ref{tab:markerword_avgs} and visually in figure \ref{fig:markerword}. Appendix \ref{app:markerword} shows the graphs for all studied words, we choose a subset of four words to show the most common trends.

Most of the studied words show a sharp increase after 2022. For example, \texttt{underscoring} and \texttt{enhancing} had relatively stable frequencies for over a decade before showing a clear trend break corresponding with the release of ChatGPT. \texttt{Underscoring} shows a ten-fold increase between 2020 and 2025 with \texttt{enhancing} showing a 2.4 increase.

In contrast, words like \texttt{leveraging} were already increasing in frequency, making it difficult to attribute their current spike solely to AI.

Although in general \texttt{delve} shows a 3.4 increase in 2025 compared to 2020, this varies between conferences. We observe that usage of the word dropped in the 2025 editions of USENIX and CCS. A possible explanation is that these deadlines occurred after the release of GPT-4.5, a more advanced model than GPT-4 \footnote{https://openai.com/index/introducing-gpt-4-5/}. It is possible that newer models influence writing styles differently or favor different vocabulary, leading to fluctuations in specific marker words.

\section{Discussion}

\subsection{Differing policies}
Despite the widespread adoption of ChatGPT in 2022, A* conferences were relatively late to adopt formal policies. NDSS led the way for its 2025 edition, with others following in 2026. The resulting landscape is fragmented, with USENIX maintaining the most lax requirements. This inconsistency can be challenging for authors submitting to multiple venues. We encourage conference organizers to converge toward standardized policies to ensure clarity and fairness across the community.

\subsection{Writing influence}
Our results show a clear rise in the usage of specific words following the introduction of GenAI. While AI is a helpful tool for structuring and polishing research, it can also introduce a vocabulary that an author might not otherwise use. It is crucial for authors to be aware of this influence and ensure that AI assistance does not compromise their unique writing style or the clarity of their results.

Additionally, we see a steady increase in lexical complexity measured in reading ease and long words usage. This can make research harder to understand over time for less-technical or less-academic readers. We believe it is crucial for authors to keep their general readability in mind as this will make their work more accessible. Future research should continue to monitor these trends as language models and community norms evolve.

\section{Conclusion}

In this paper, we have presented a longitudinal analysis of twenty-five years of cybersecurity research, tracing the evolution of writing styles and the recent impact of Large Language Models. We observed a measurable increase in lexical complexity characterized by higher long-word rates and decreased readability
Additionally, we find a sharp rise in the frequency of LLM-favored marker words such as \texttt{underscoring} and \texttt{enhancing}.

Furthermore, our examination of conference policies reveals a fragmented landscape. While the community has begun to acknowledge the role of GenAI, the requirements for disclosure and the definitions of acceptable use vary across venues. This lack of standardization may create uncertainty for authors.

As generative models continue to evolve, it is likely that the stylistic signatures we identified will continue to shift. We conclude that while LLMs offer valuable assistance in overcoming language barriers and improving the efficiency of the writing process, the community must remain vigilant. Authors should ensure that the usage of AI-assisted writing does not come at the expense of harder-to-read text.

\section{LLM Usage}
With the rise of LLMs, the question frequently asked is how useful these models can be when leveraged for more than a writing assistant.
In order to test and showcase the capabilities of current models, the research and writing in this paper, aside from this section, was heavily performed using various generative AI models. All output was manually verified for its correctness.
To ensure transparency and because it can benefit the community, this section explains how we instructed the various LLMs to perform research and the lessons learned from performing research with LLMs.
The used prompts can be found in appendix \ref{app:prompts}.
We hope our insights can lead to further discussions on how the community should handle AI-generated papers containing interesting scientific results, where humans checked for correctness, yet are not the main driver behind the idea.

\subsection{Instructions}
As a general research assistant tasked with all analytical tasks we used GPT5.2 \footnote{https://openai.com/index/introducing-gpt-5-2/} for its deep thinking qualities. It was given a custom instruction to behave as a research assistant together with some extra details to prevent hallucination of references (Prompt \ref{prompt:general}, partially inspired by Claude guardrails \footnote{https://platform.claude.com/docs/en/test-and-evaluate/strengthen-guardrails/reduce-hallucinations}).
To bootstrap the research project, we presented the model with our general idea: the change of language over time in cybersecurity papers focusing on the influence of LLMs. Based on this general idea, it was tasked to search for sources online and identify gaps in this topic in the current literature to come up with concrete research questions (Prompt \ref{prompt:start}).

GPT5.2 identified 8 research questions of which we kept three and asked the model to further refine them whilst generating Python code to perform the analysis (Prompt \ref{prompt:code}).
To identify the conference policies, we manually created the table highlighting the main differences.
All code was further refined to match our dataset structure and fix certain logical errors using Claude Sonnet 4.5 integrated as GitHub Copilot in our editor.

After we populated our dataset with the linguistic markers using the generated scripts, we tasked the LLM to create a script to analyze the results and output the results in a way it can later interpret them to figure out its conclusion (Prompt \ref{prompt:analyse}).
GPT5.2 provided instructions on how to return the analysis script and what data it needed as input for its next query.
Feeding the analyzed data back into the model provided us with 6 "key insights" of which we picked 3 to generate graphs for.
To make sure the generated figures are correct, we always asked for the code to create the graphs ourselves.

After all the results and graphs were obtained the LLM generated text, each in a separate prompt, for the following sections: introduction, background \& related work and the results. 
To further refine the text, we first asked GPT5.2 in a new conversation to streamline the storyline of the paper whilst also feeding it previous publications of the authors to base its style and tone on.
Finally, the discussion section was written manually while the abstract and conclusion were generated each in a new chat without previous context aside from the current draft.

The final LLM usage existed of feeding the draft to Gemini 3 which, in 2 separate conversations, first refined the text with grammar, spell and style checked and afterwards was tasked to behave as a reviewer made aware of the submission target. 
Based on the review, we manually altered certain sections and sentences finishing the paper.

\subsection{Lessons Learned}
While performing the research presented in this paper whilst heavily relying on generative AI, some hurdles were encountered which are formulated as advice below

\textbf{Use multiple chats} After generating lots of files and images, we noticed a decrease in response quality likely due to the context window being reached or the LLM loosing track of its original purpose. Therefore, we recommend splitting up different tasks in different chats. For example, one chat could be used to analyze results and interpret them, one as a background checker fed with, manually, collected references and a final chat for writing.

\textbf{Use multiple models} To limit the impact a single model with its own limitations can have on a paper, it is recommended to use a second model for a final rewrite or to act as a reviewer. This will give more insights, possibly from different perspectives compared to using a single model.

\textbf{Writing assistant} We showed in this paper that generative AI influences the writing style of which authors need to be aware. Therefore, it can be helpful to feed previous work to the model first to mimic its style for future text generation. Generated text should afterwards be checked carefully to avoid plagiarism as we noticed the model can copy full sentences if they fit the narrative.
Additionally, if not specifically asked to perform a spell check, we noticed the LLMs often copied over our grammatical and spelling mistakes without changing them.

\bibliographystyle{IEEEtran}
\bibliography{references}

\appendices
\section{Marker word graph}
\label{app:markerword}
\begin{figure*}
  \centering
  \includegraphics[width=0.95\textwidth]{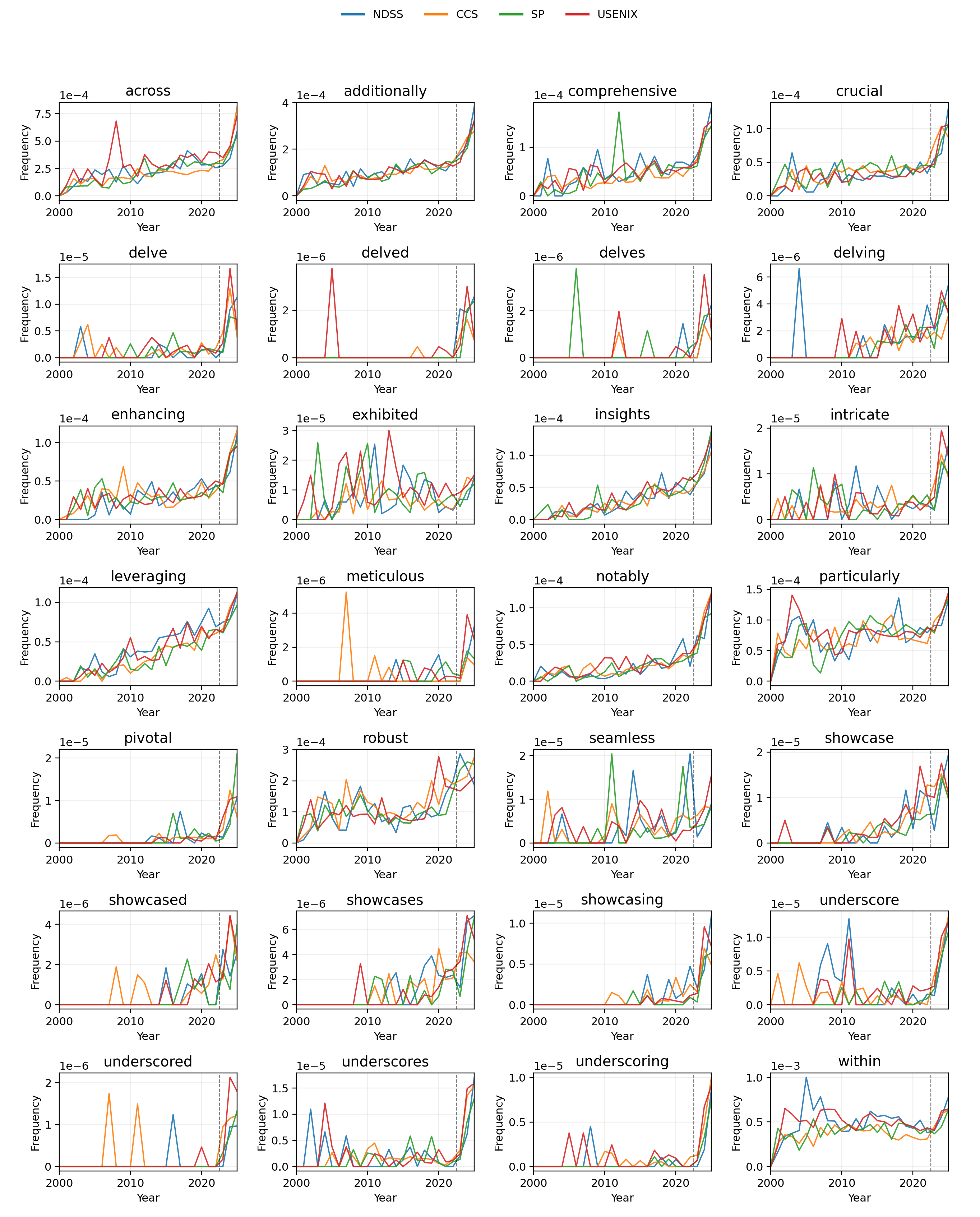}
  \caption{Full set of tracked marker words showing longitudinal frequency trends.}
  \label{fig:markerwordall}
\end{figure*}

\input{prompts.tex}
\end{document}

%% file: policytable.tex
\begin{table*}[th]
    \caption{Overview of Generative AI policies showing: if its specifically noted that authors bear full responsibility; If a dedicated section declaring AI usage is mandatory; Whether using GenAI for grammar and spell checks can be used without mentioning it; If authors need to address enviromental concerns when using GenAI.}
    \label{tab:policy}
    \centering
    \begin{tabular}{lcccc}
        \textbf{Conference} & \textbf{Author Responsibility} & \textbf{Dedicated Section} & \textbf{Grammar-check} & \textbf{Enviromental Implications} \\
        \midrule
        NDSS '25 & \xyes & \xhalf & \xyes & \xno\\
        NDSS '26 & \xyes & \xhalf & \xyes & \xno\\
        \midrule
        USENIX '26 & \xyes & \xno & \xyes & \xno\\
        \midrule
        CCS '26 & \xyes & \xyes & \xyes & \xno \\
        \midrule
        SP '26 & \xyes & \xyes & \xno & \xyes \\ 
    \end{tabular}
\end{table*}

%% file: prompts.tex
\section{Used Prompts}
\label{app:prompts}

\begin{lstlisting}[basicstyle=\small, caption={Custom instructions set on all GPT5.2 chats.}, captionpos=b, label=prompt:general]
You are an expert research assistant tasked
with aiding projects to be published
in top-tier cybersecurity venues.
When unsure about any aspect of the task or
if you lack information please state
the information needed and answer on that
part with "I don't have enough information
to confidently assess this."

For each claim you make that is not grounded
in the experiments conducted in the project
you're currently working on, find a document
or scientific paper that supports said claim.
If no support can be found,
mark that claim with empty [] brackets.
\end{lstlisting}

\begin{lstlisting}[basicstyle=\small, caption={First prompt used to identify the research questions based on existing literature.}, captionpos=b, label=prompt:start]
The project you will be working on will
study the change of language over time in
top-tier cybersecurity papers focusing on the
influence LLM might have had on writing style.
Search existing research on the topic and
identify gaps in the current research.
Based on these gaps come up with
adequate research questions.

In a later stage of the project,
you will conduct experiments to answer the research
questions on a dataset of all papers published
in A* conferences in cybersecurity of the past 25 years.

Keep track of all the sources you used.
\end{lstlisting}

\begin{lstlisting}[basicstyle=\small, caption={General prompt used to generate the scripts for the analysis per research question.}, captionpos=b, label=prompt:code]
We now focus on the code needed for the analysis.
First we focus on RQ1. Come up with good metrics,
based on previous research, that calculate lexical,
syntactic, readability and/or rhetorical features.
Create a .py file that calculates these metrics.
\end{lstlisting}

\begin{lstlisting}[basicstyle=\small, caption={Prompt to make the model analyze the results.}, captionpos=b, label=prompt:analyse]
Given all the results from RQ1, 2 and 3 in the mongoDB.
Generate a script to analyze the results.
When providing you with the output of said script
you will need to select the most interesting results
to further analyze and later explain in the paper.
\end{lstlisting}